\documentclass[fleqn,showpacs,showkeys,preprintnumbers,twocolumn,
nofootinbib]{revtex4-1}
\usepackage{graphicx}
\usepackage{amssymb}
\usepackage{amsmath}
\usepackage{bm}

\begin{document}
\preprint{INHA-NTG-01/2013}
\title{Transverse charge densities in the nucleon in nuclear matter}

\author{Ulugbek Yakhshiev}
\email{yakhshiev@inha.ac.kr}
\affiliation{Department of Physics, Inha University, Incheon
402-751, Republic of Korea}

\author{Hyun-Chul Kim}
\email{hchkim@inha.ac.kr}
\affiliation{Department of Physics, Inha University, Incheon
402-751, Republic of Korea}

\date{April 2013}

\begin{abstract}
We investigated the transverse charge densities in the nucleon in
nuclear matter within the framework of the in-medium modified Skyrme model.
The medium modification of the nucleon electromagnetic form factors
are first discussed. The results show that the form factors in nuclear
matter fall off faster than those in free space, as the
momentum transfer increases. As a result, the charge radii of the
nucleon become larger, as the nuclear matter density increases.
The transverse charge densities in the nucleon indicate that
the size of the nucleon tends to bulge out in nuclear matter. This
salient feature of the swelling is more clearly observed in the
neutron case. When the proton is transversely polarized, the
transverse charge densities exhibit the distortion due to the
effects of the magnetization.
\end{abstract}


%12.39.Dc Skyrmions
%13.40.Gp Electromagnetic form factors
%14.20.Dh Protons and neutrons


\maketitle

%\section{Introduction}
\textbf{1.} It is of utmost importance to understand the
structure of the nucleon in particle and nuclear physics, since the
nucleon consists of the basic building block of matter. In
particular, the nucleon electromagnetic (EM) form factors are the
fundamental issue in that they reveal how the electric charge and
magnetization of the quarks are distributed inside the nucleon. While
the EM form factors have been studied well over several
decades, their understanding is still not complete.

Recently, a series of new measurements of the EM form factors has been
carried out and has produced the remarkably precise
data~\cite{Jones:1999rz, Gayou:2001qt,
  Gayou:2001qd,Punjabi:2005wq,Puckett:2010ac, Bernauer:2010wm,
  Ron:2011rd, Zhan:2011ji}. These new experimental
results have subsequently intrigued both experimental and theoretical
works (see, for example, the following
reviews~\cite{HydeWright:2004gh,Arrington:2006zm,Perdrisat:2006hj}).
The experimental data with high precision enable one to make a flavor
decomposition of the nucleon EM form factors with isospin and charge
symmetries taken into account~\cite{Cates:2011pz,Qattan:2012zf}.
Moreover, generalized parton distributions (GPDs) have unveiled a
novel aspect of the nucleon EM structure: The Fourier transforms of
the nucleon EM form factors in the transverse plane, as viewed from a
light front frame moving towards a nucleon, paint a tomographic
picture of how the charge densities of quarks are
distributed~\cite{Burkardt:2002hr,Burkardt:2000za} transversely.
These transverse charge densities of the quarks inside a nucleon have
been already investigated for the unpolarized~\cite{Miller:2007uy} and
transversely polarized~\cite{Carlson:2007xd} nucleons.

It is of equal importance to examine
how the EM structure of the nucleon is changed in nuclear
matter. Studying the EM form factors of the nucleon in medium
provides a new aspect on EM properties of the nucleon modified in
nuclei. In fact, the first experimental study of deeply virtual
Compton scattering on (gaseous) nuclear targets (H, He, N, Ne, Kr, Xe)
was reported in~\cite{Airapetian:2009bi}. While uncertainties of
the first measurement are so large that one is not able to observe
nuclear modifications of the nucleon structure, Future experiments
will bring about more information on medium modifications of the EM
properties of the nucleon.

In the present work, we want to investigate the nucleon EM form
factors and the transverse charge densities of quarks inside a
nucleon in nuclear matter within the framework of an in-medium
modified Skyrme model. The Skyrme model has certain virtues: it is
simple but respects chiral symmetry and its spontaneous symmetry
breaking. Moreover, one can easily extend it to the study of nuclear
matter, based on modifications of the pion in
medium~\cite{Rakhimov:1996vq,Yakhshiev:2010kf}. The energy-momentum tensor form
factors of the nucleon, which are yet another fundamental form factors
that are related to the generalized EM form factors, have been
investigated in nuclear matter within this
framework~\cite{Kim:2012ts}. The results have explained certain
interesting features of the modifications of the nucleon in nuclear
matter such as the pressure and angular momentum.
Indeed, we will also show in this work how
the EM properties of the nucleon are changed in nuclear matter in a
simple manner. We will also see that the transverse charge densities
expose noticeably how the distribution of quarks undergo
changes in the presence of nuclear medium.

\vspace{0.5cm}
%\section{Effective chiral Lagrangian}
\textbf{2.} We begin with the in-medium modified effective chiral
Lagrangian~\cite{Yakhshiev:2010kf}:
\begin{eqnarray}
\mathcal{L}^*&=&\frac{F_\pi^2}{16}\,{\rm Tr}\left(\frac{\partial
U}{\partial t}\right)\left(\frac{\partial U^\dagger}{\partial
t}\right) \cr
&&-\frac{F_\pi^2}{16}\,\alpha_p({\bm r}){\rm
  Tr}({\bm\nabla} U)\cdot({\bm\nabla}
 U^\dagger)\cr
&&+\frac{1}{32e^2\gamma({\bm r})}\,{\rm
Tr}[U^\dagger\partial_\mu U,U^\dagger\partial_\nu U]^2\cr
&&+\frac{F_\pi^2m_\pi^2}{16}\,\alpha_s({\bm r}){\rm Tr}(U+
U^\dagger-2)\,, \label{Eq:Lag}
\end{eqnarray}
where $F_\pi=108.78$~MeV denotes the pion decay constant, $e=4.85$
the Skyrme parameter, and $m_\pi$ the experimental value of
the pion mass $m_{\pi^0}=134.98$~MeV. This set of the parameters
reproduce qualitatively well the experimental data for the nucleon
and $\Delta$-isobar in free space.

The medium functionals, $\alpha_s$, $\alpha_p$ and $\gamma$, are
expressed as
\begin{eqnarray}
\alpha_s&=&1-\frac{4\pi b_0\rho({\bm r})f}{m_\pi^{2}},\cr
\alpha_p&=&1-\frac{4\pi c_0\rho({\bm r})}{f+g_0'4\pi
c_0\rho({\bm r})},\cr
\gamma&=&\exp\left(-\frac{\gamma_{\mathrm{num}} \rho({\bm
      r})}{1+\gamma_{\mathrm{den}}\rho({\bm r})}\right)\,.
\label{medfunc}
\end{eqnarray}
They encode information on how the surrounding environment influences
properties of the single skyrmion. The parameters $\alpha_s$ and
$\alpha_p$ are related to the corresponding phenomenological $S$- and
$P$-wave pion-nucleus scattering lengths and volumes,
i.e. $b_0=-0.024m_\pi^{-1}$ and  $c_0=0.06m_\pi^{-3}$,
respectively, which describe the pion physics in a
nucleus~\cite{Ericsonbook}. The last functional $\gamma$
is parameterized in the form of an exponential function and represents
the medium modification of the Skyrme parameter. This simple form with
two variational parameters $\gamma_{\rm num}=0.47m_\pi^{-3}$ and
$\gamma_{\rm den}=0.17m_\pi^{-3}$ reproduce the correct position of
the saturation point of symmetric nuclear
matter~\cite{Yakhshiev:2010kf}. The $\rho$ stands for the density of
nuclear matter. The $g_0^{\prime}=0.7$ denotes the Lorentz-Lorenz or
correlation parameter and $f=1+m_\pi/m_{N}^{\rm free}$ is the
kinematical factor.

Since we consider isospin symmetric infinite nuclear matter in the
present work, the density can be regarded as a homogeneous and
constant one. Thus, we can simply choose the spherically symmetric
``hedgehog" for the soliton:
\begin{equation}
U=\exp\{i\,{\bm n}\cdot{\bm \tau}\,F(r)\},
\end{equation}
where ${\bm n}$ denotes the unit radial vector in coordinate space and
${\bm \tau}$ the usual Pauli isospin matrices.

\vspace{0.5cm}
%\section{Electromagnetic properties of the bound nucleons}
\textbf{3.} We refer to Ref.~\cite{Yakhshiev:2010kf} for the details
of the minimization procedure and other useful formulas. In this work,
we concentrate on EM properties of the nucleon.

The nucleon matrix element of the EM current is expressed in
terms of the Dirac and Pauli form factors:
\begin{eqnarray}
&& \hspace{-1cm}\langle N(p',\,S') | J_\mu^{EM} (0) |N(p,\,S)\rangle   \cr
&& \hspace{-1cm} =\overline{u}_N(p',\,S') \left[
\gamma_\mu F_1^*(q^2) +i \frac{ \sigma_{\mu\nu}q^\nu}{2m_N}
  F_2^*(q^2) \right] u_N(p,\,S),
\end{eqnarray}
where the EM current is defined in terms of the baryon current $B_\mu$
and the isovector current $J_\mu^{(3)}$
\begin{equation}
J_\mu^{EM} (0) \;=\; \frac12 (B_\mu(0) + J_\mu^{(3)}(0)).
\end{equation}
The $\gamma_\mu$ denotes the Dirac matrices and $u_N(p,\,S)$ stands for
the Dirac spinor for the nucleon with mass $m_N$, momentum $p$, and
the third component of its spin $S$. The $\sigma_{\mu\nu}$ is the spin
operator $i[\gamma_\mu,\,\gamma_\nu]/2$ and $q^2$ the square of the
momentum transfer $q^2=-Q^2$ with $Q^2>0$. The asterisk means the form
factors in medium. In the Breit frame, the in-medium modified Sachs EM
form factors of the nucleon are
defined as
\begin{eqnarray}
G_{E}^*(Q^2)&=&\frac12\int{\rm d}^3r\,e^{i\bm q\cdot \bm r}
J_0^{EM} ({\bm r})\,,\nonumber\\
G_{M}^*(Q^2)&=&\frac{m_N}{2} \int{\rm d}^3r\,e^{i
  \bm q\cdot \bm r}[{\bm r}\times {\bm J}^{EM}({\bm r})]\,,
\end{eqnarray}
where $J^0$ and $\bm J$ correspond respectively to the temporal and
spatial components of the properly normalized sum of the baryonic
(topological) current $B_\mu$ and the third component of the isovector
(Noether) current ${V}_\mu^*$.

The isoscalar ($S$) and isovector
EM formfactors are generically expressed as
\begin{eqnarray}
G_{E,\,M}^{S,\,V,*}(Q^2)&=&\int {\rm d}^3r\,
e^{iqr\cos\theta}\rho_{E,\,M}^{S,\,V}(r,\theta),
\label{ff}
\end{eqnarray}
where the densities are given as
\begin{eqnarray}
\rho_{\rm E}^{\rm S}(r,\theta)&=&-\frac{F' \sin^2F }{4\pi^2r^2}\,,\cr
\rho_{\rm M}^{\rm S}(r,\theta)&=&-\frac{m_N}{8\pi^2\lambda^*}\,
F' \sin^2F \sin^2\theta\,, \cr
\rho_{\rm E}^{\rm V}(r,\theta)&=&\frac{\sin^2F}{12\lambda^*}
\left\{F_\pi^2+
  \frac{4}{e^2\gamma}\left(F_r^2+\frac{\sin^2F}{r^2}\right)\right\}\,,
\cr
\rho_{\rm M}^{\rm V}(r,\theta)&=&\frac{m_N}{3}\left\{F_\pi^2\alpha_p+
\frac{4}{e^2\gamma}\left(F'^2+\frac{\sin^2F}{r^2}\right)\right\} \cr
&& \times \sin^2F \sin^2\theta\,.
\label{chargedists}
\end{eqnarray}
Here, $\lambda^*$ stands for the medium-modified moment of inertia of
the soliton. Note that because of the modification of the Skyrme term,
all of the electromagnetic charge distributions except for the
isoscalar electric charge depend explicitly on the medium
density. Moreover, the medium functional $\alpha_p$ appears in the
expression of the isovector magnetic density distribution.

The charges of the proton ($p$) and neutron ($n$) are defined as
\begin{eqnarray}
\left(
  \begin{array}{c}
Q^p \\ Q^n
  \end{array}
\right)
%&=&\frac{B}{2}+T_3^{\left(\rm p\atop n\right)}
&:=& \int {\rm d}^3r\,\left\{\rho_{E}^{S}(r,\theta)\pm
\rho_{E}^{V}(r,\theta)\right\}\,,
\label{charge}
\end{eqnarray}
where density distributions include the prefactor $1/2$. Similarly,
the magnetic moments of the nucleon are defined as
\begin{eqnarray}
\left(
  \begin{array}{c}
\mu_p \\ \mu_n
  \end{array}
\right)
&:=&
\int {\rm d}^3r\left\{\rho_{ M}^{ S}(r,\theta)\pm
\rho_{ M}^{ V}(r,\theta)\right\}\,.
\end{eqnarray}
The corresponding isoscalar and isovector mean square charge radii are
given as
\begin{eqnarray}
\langle r^2\rangle_{I=0}^*&=&-\frac2\pi\int_0^\infty F'\sin^2 F\,r^2
{\rm d}r\,,\\
 \langle r^2\rangle_{I=1}^* &=&\frac{2\pi}{3\lambda^*}\int_0^\infty
\left[F_\pi^2+\frac4{e^2\gamma}\left(F'^2+\frac{\sin^2F}{r^2}\right)\right]
\cr
&&
\times \sin^2 F r^4 {\rm d}r\,.\nonumber
\end{eqnarray}

\vspace{0.5cm}
%\section{Results and discussions}
\textbf{4.} We are now in a position to discuss the results. In
Table~\ref{tab1}, we list the results of the magnetic moments,
the isoscalar and isovector charge radii of the nucleon as functions
of $\rho/\rho_0$. The electric charge radius of the nucleon is well
reproduced in free space, compared with the experimental data:
$(0.877\pm 0.005)\,\mathrm{fm}$.
\begin{table}[htb]
\begin{ruledtabular}
\begin{tabular}{ccccc}\noalign{\smallskip}
$\rho/\rho_0$ &$\mu_{p}^*$ [n.m.]&$\mu_{n}^*$ [n.m.]&$\sqrt{\langle
r^{2}_{p}\rangle^{*}}$ [fm]& $\langle r^{2}_{n}\rangle^{*}$
[fm$^2$]\\
\noalign{\smallskip}\hline\noalign{\smallskip}
       0.0 &      1.965  &    $-1.238$   &    0.882   &   $ -0.316 $\\
       0.5 &      2.134  &    $-1.430$   &    0.926   &   $ -0.310 $\\
       1.0 &      2.315  &    $-1.634$   &    0.966   &   $ -0.304 $\\
\end{tabular}
\end{ruledtabular}
\caption{Magnetic moments, meansquare radii of
the nucleons in nuclear matter. Magnetic moments are given in units of
nuclear magnetons (n.m.). }
\label{tab1}
\end{table}
The results of the proton charge radius shows that the size of the
proton swells up in nuclear matter, which is consistent with those of
different approaches. For example, at normal nuclear matter density
the proton charge radius $\sqrt{\langle r_p^{2}\rangle^*}$
is enhanced  by about $10\,\%$. On the other hand, the magnitude of
the neutron charge radius $\langle r_n^{2}\rangle^*$ is reduced by
about $4\,\%$. At first glance, the size of the neutron seems
reduced. In fact, however, it swells like that of the proton, because
the positive charges of the neutron are distributed in the inner part
of, whereas the negative charges are found in outer part. When the
neutron is in medium, both the positive and negative charges are
broadened, which implies that size of the neutron actually becomes
larger in nuclear matter. This will be explicitly shown later when we
discuss the transverse charge densities of the neutron.

As was known already, the results of the nucleon
magnetic moments in the Skyrme model turn out to be
underestimated. However, since we are mainly interested in the
modification of the nucleon magnetic moments in nuclear matter, we
will concentrate on how the magnetization of the nucleon undergoes
changes in medium. The magnetic moments are even more dramatically
changed due to the additional medium factor $\alpha_p$ in the
expression of the magnetic-vector density distribution
$\rho_{ M}^{V}$.
As a result, the proton magnetic moment is enhanced by almost
about $18\,\%$ at normal nuclear matter density $\rho_0$, whereas the
neutron one increases by about $30\,\%$. It indicates that the neutron
is changed more sensitively, which will be shown later again in the
neutron transverse charge densities.

\begin{figure*}[ht]
\centerline{\resizebox{0.9\columnwidth}{!}{\includegraphics{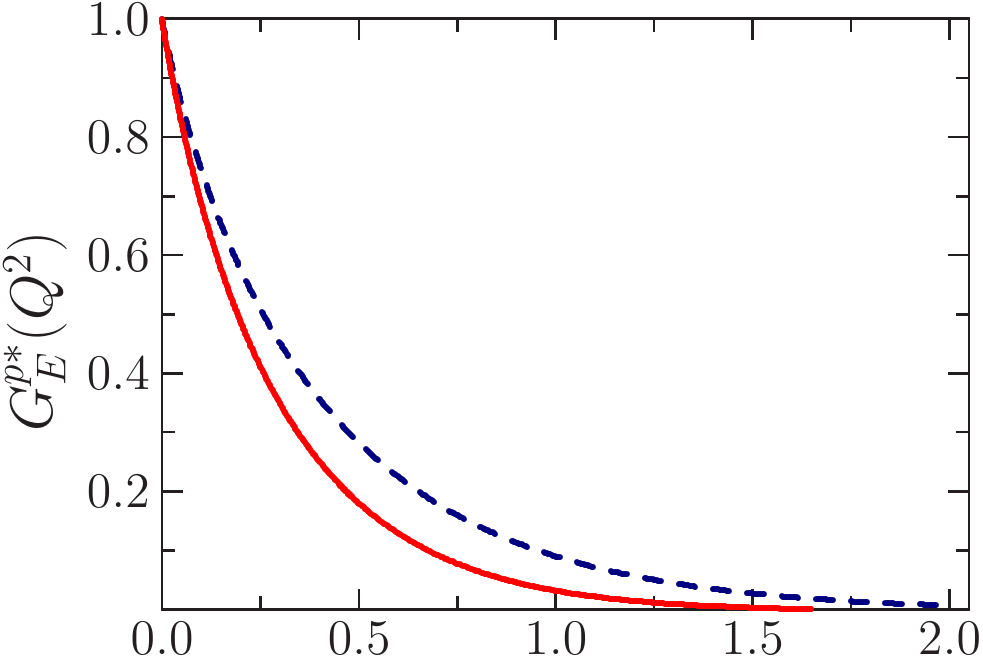}}
\hskip 0.6cm\resizebox{0.9\columnwidth}{!}{\includegraphics{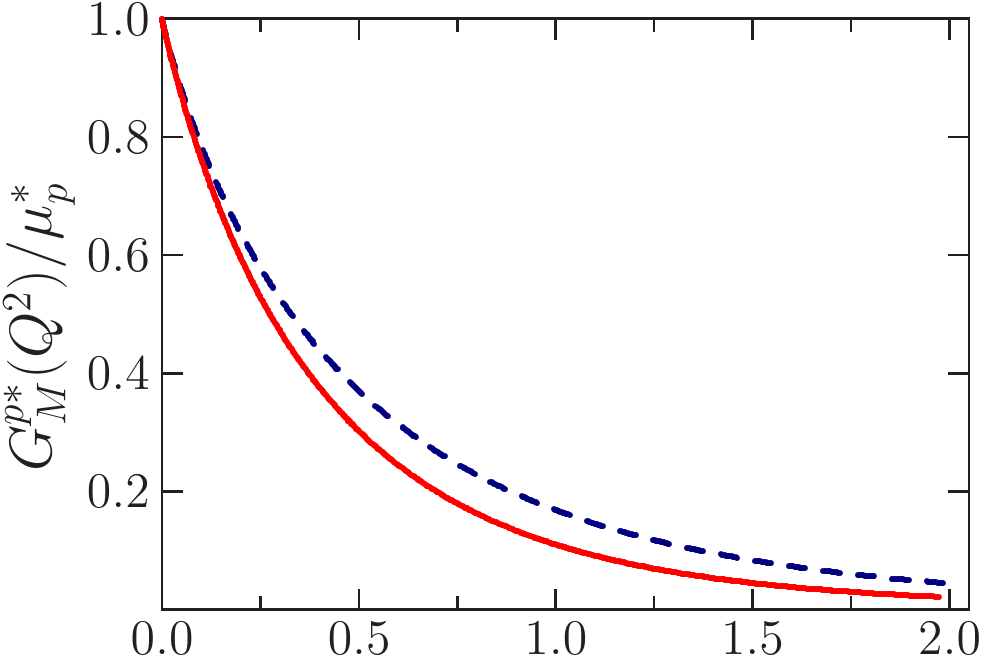}}}
\vskip 0.5 cm
\centerline{\resizebox{0.92\columnwidth}{!}{\includegraphics{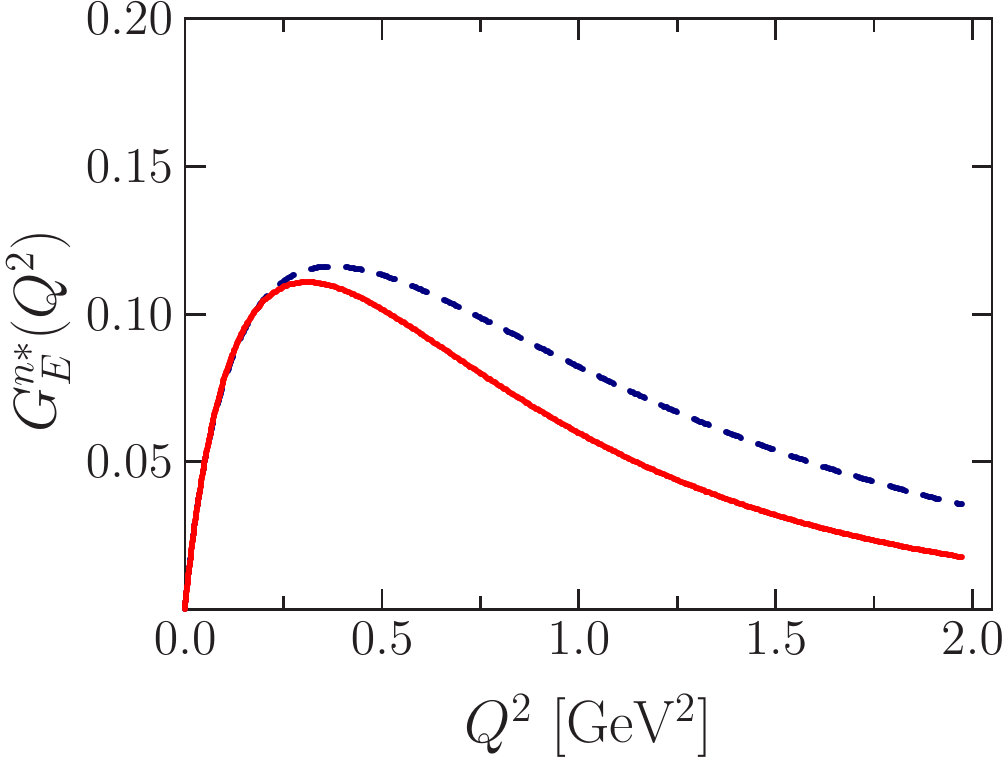}}
\hskip 0.6cm\resizebox{0.9\columnwidth}{!}{\includegraphics{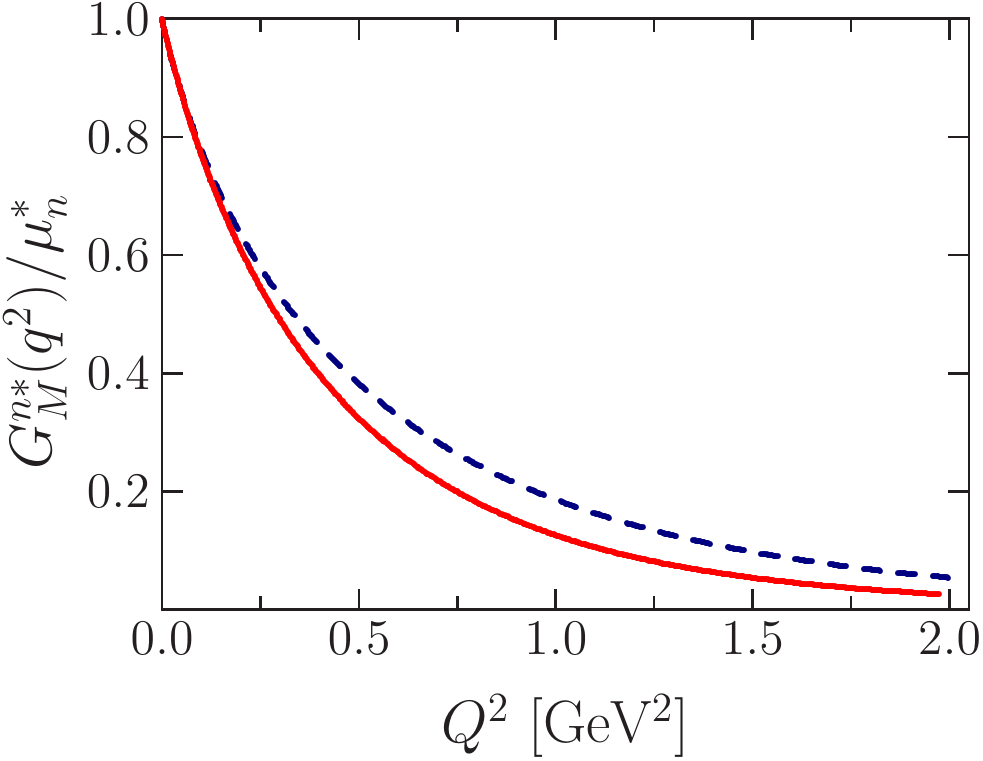}}}
\caption{(Color on line) The electric (left panels) and the normalized
  magnetic (right
  panels) form factors of the nucleons  as a function of squared
  momentum transfer $Q^2$. The dashed curve corresponds to the
free space case, the solid one in nuclear matter with the
normal nuclear matter density  $\rho_0=0.5m_\pi^3$. }
\label{fig:pnEMff}
\end{figure*}
Figure~\ref{fig:pnEMff} draws the results of the proton
and the neutron EM form factors in free space (dashed curve)
and at normal nuclear matter density $\rho_0$ in nuclear matter
(solid curve), respectively. We observe that as the density
increases, the EM form factors fall off in general faster than those
in free space. Similar qualitative results have been obtained in the
framework of the quark meson coupling (QMC)
model~\cite{Lu:1997mu}. However, the present results
are quantitatively different from those in the QMC model. For
example, compared with the proton electric form factor at
$Q^2=0.3\,\mathrm{GeV}^2$ in free space, $G_E^{p*}(Q^2)$ is reduced by
about $30\,\%$ at $\rho_0$, whereas that from the QMC model
decreases only by about $8\,\%$~\cite{Lu:1997mu}. In the
case of the neutron one $G_{E}^{n*}(Q^2)$, it is the other way
around. At $Q^2=0.3$~GeV$^2$, the neutron electric form factor is
changed by about $1\,\%$ only but in the QMC model brings it is
lessened by about $8\,\%$. However, the neutron electric form factor
falls off faster in nuclear matter than that in free space (e.g. at
$Q^2=1$~GeV$^2$ $G_E^{n*}$ is decreased by approximately
$30\%$).  On the other hand, the present results are similar to those
of Ref.~\cite{Meissner:1988wj}. In general, the neutron is
more likely to be changed in medium.

The modifications of the nucleon magnetic form factors
are also more prominent within the present model in
comparison with the QMC model. At normal nuclear
matter $\rho=\rho_0$ and at $Q^2=0.3\,\mathrm{GeV}^2$ the proton and
neutron magnetic form factors are reduced by about $12\,\%$ and
$8,\%$, respectively, compared to the case of the free space
within our approach, while the QMC model yields them
decreased by $1.5\,\%$ and $0.9\,\%$, respectively. Thus, in general,
the present results show stronger medium effects than those from the
QMC model.

\begin{figure*}[hbt]
\centerline{\resizebox{0.9\columnwidth}{!}{\includegraphics{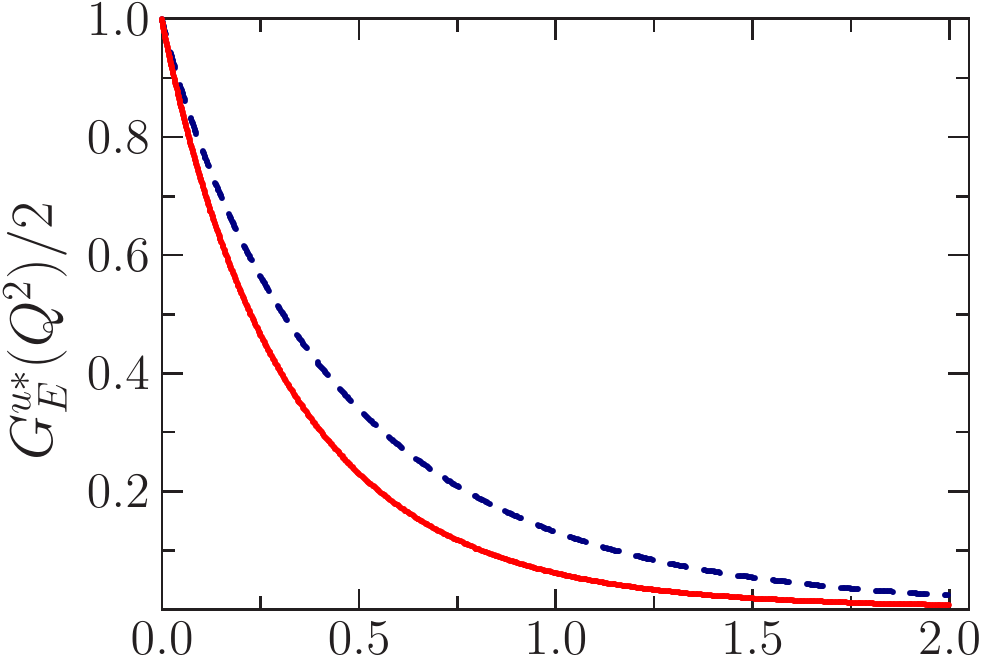}}
\hskip 0.6cm \resizebox{0.9\columnwidth}{!}{\includegraphics{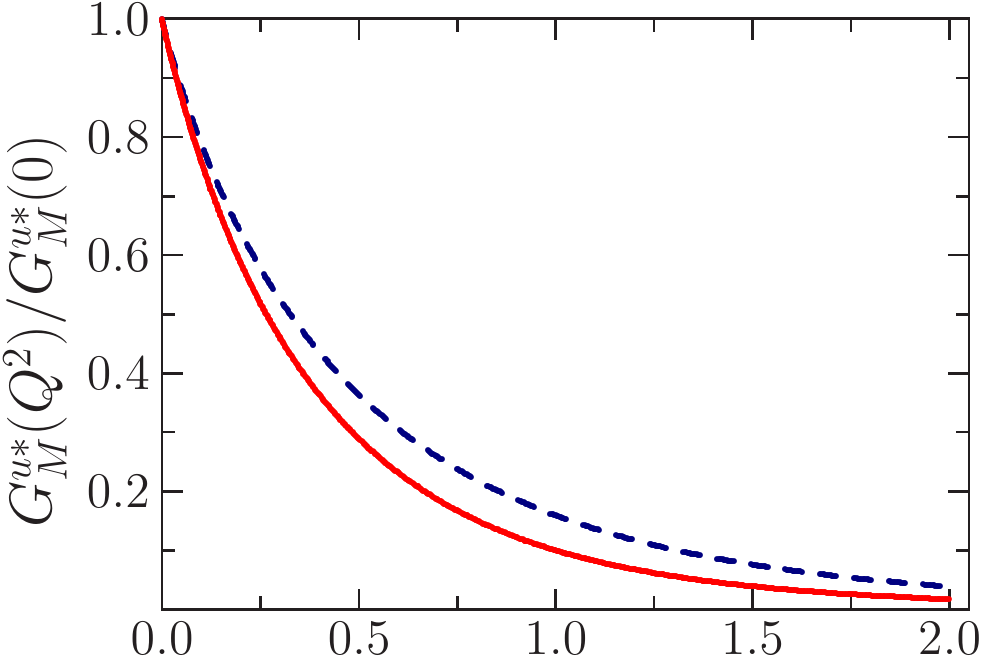}}}
\vskip 0.5cm
\centerline{\resizebox{0.9\columnwidth}{!}{\includegraphics{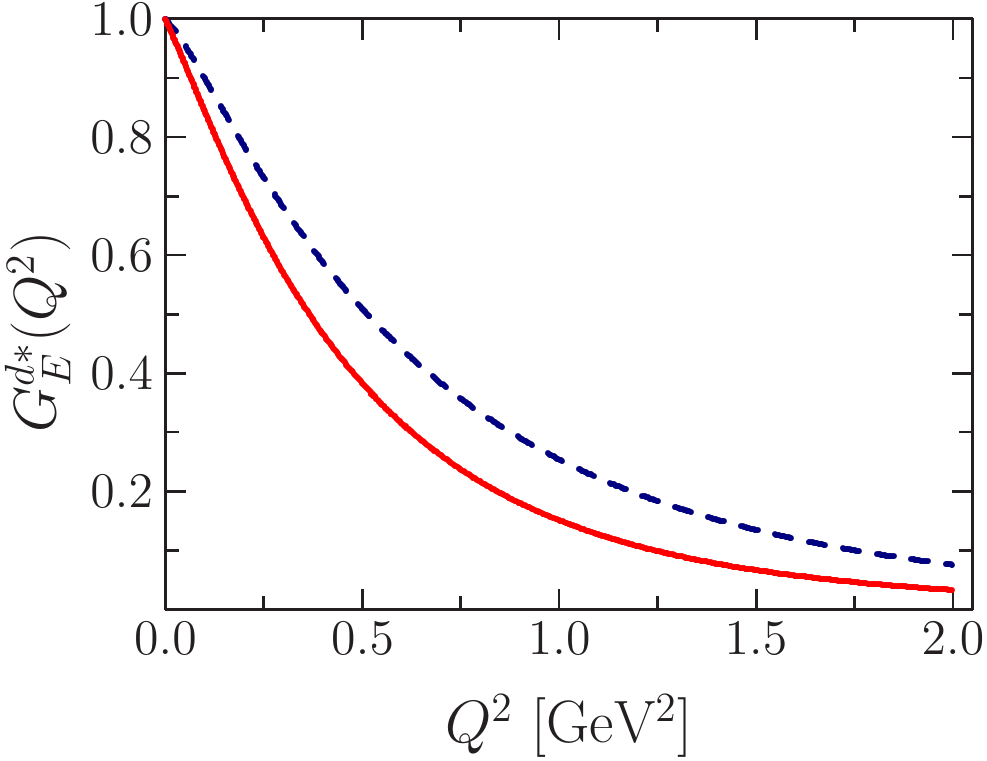}}
\hskip 0.6cm \resizebox{0.9\columnwidth}{!}{\includegraphics{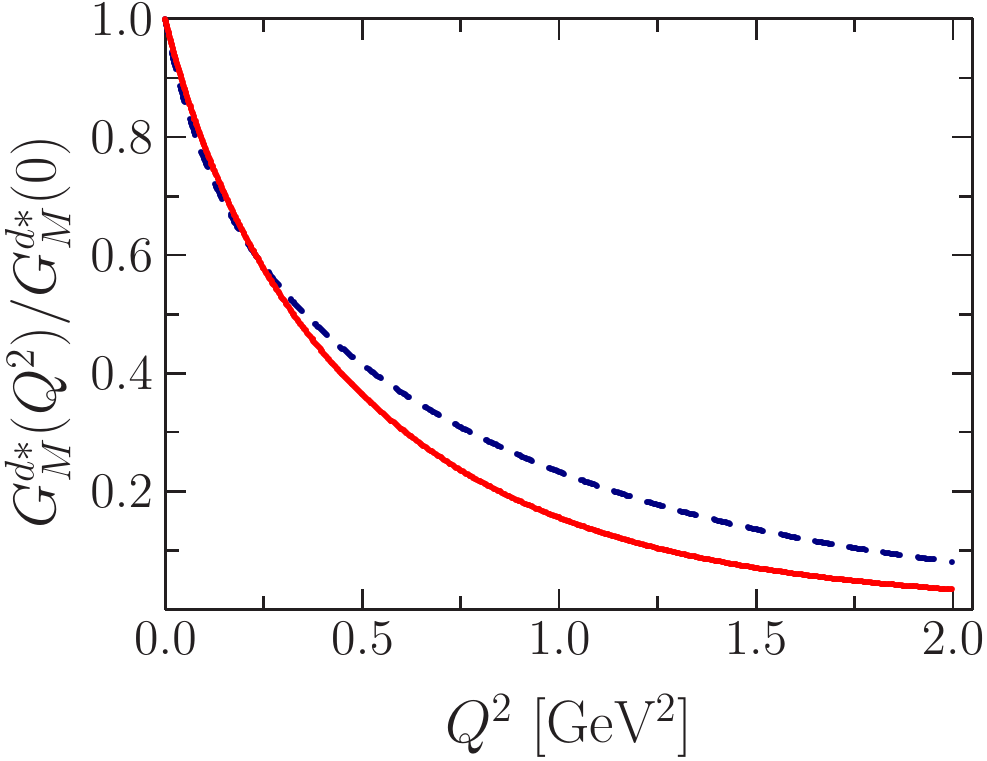}}}
\caption{The electric (left panels) and normalized magnetic (right panels)
form factors of the up (upper panels) and down (lower panels) quarks
inside a nucleon in nuclear matter as a function of squared momentum
transfer $Q^2$. Other notations are the same as in
Figure~\ref{fig:pnEMff}. }
\label{fig:udff}
\end{figure*}

It is also of great interest to study the flavor-decomposed EM form
factors, since the experimental data are now available in free
space~\cite{Cates:2011pz,Qattan:2012zf}. We want to examine how the
flavor-decomposed EM form factors undergo the changes in nuclear
matter. Assuming isospin symmetry in flavor SU(2), we can decompose
the nucleon EM form factors into those of the up and down quarks as
follows~\cite{Miller:1990iz,Beck:2001yx}:
\begin{eqnarray}
G_{E,M}^{ u}=2G_{E,M}^{ p}+G_{E,M}^{ n}\,,\cr
G_{E,M}^{ d}=G_{E,M}^{ p}+2G_{E,M}^{ n}\,.
\end{eqnarray}
In Fig.~\ref{fig:udff}, we draw the results of the up and down EM form
factors in free space (dashed curve) and at normal nuclear matter
density (solid curve). In general, the form factors in nuclear matter
tend to fall off faster than those in free space. The down EM form
factors are shown to be more sensitively affected by the presence of
nuclear matter. Interestingly, the in-medium down magnetic form factor
starts to decrease less than that in free space till about $Q^2\approx
0.3\,\mathrm{GeV}^2$, and then drops off sharply. We will soon see a
similar feature when the transverse charge densities are considered.

The transverse charge densities of quarks inside a nucleon reveal the
modification of the nucleon EM structures more prominently. In fact,
the usual charge and magnetization densities obtained from the
three-dimensional Fourier transform of the EM form factors in the
Breit framework obscure their physical meaning because of the Lorentz
contraction of the nucleon in its moving
direction~\cite{Kelly:2002if,Burkardt:2002hr}.  On the other hand, the
transverse charge density provides a clear understanding of how the
charge of quarks inside a nucleon are distributed. We can extend
straightforwardly this to nuclear matter. However, We want to mention
a caveat. Since the original Skyrme model produces the underestimated
magnetic form factors of the nucleon but explains reasonable $Q^2$
dependence, whereas it yields the overestimated electric one of the
neutron. It causes a difficulty in describing the transverse charge
density inside a neutron that depend on both the neutron electric and
magnetic form factors. However, the present aim is at studying how the
EM properties of the nucleon undergo changes in nuclear matter, so
that we will rather use the normalized nucleon magnetic form factors
with the experimental data for the corresponding magnetic moments as
shown in Fig.~\ref{fig:pnEMff}.

\begin{figure*}[hbt]
%            \parbox{\halftext}{%   %\def\halftext{.471\textwidth}
\centerline{\resizebox{0.9\columnwidth}{!}{\includegraphics{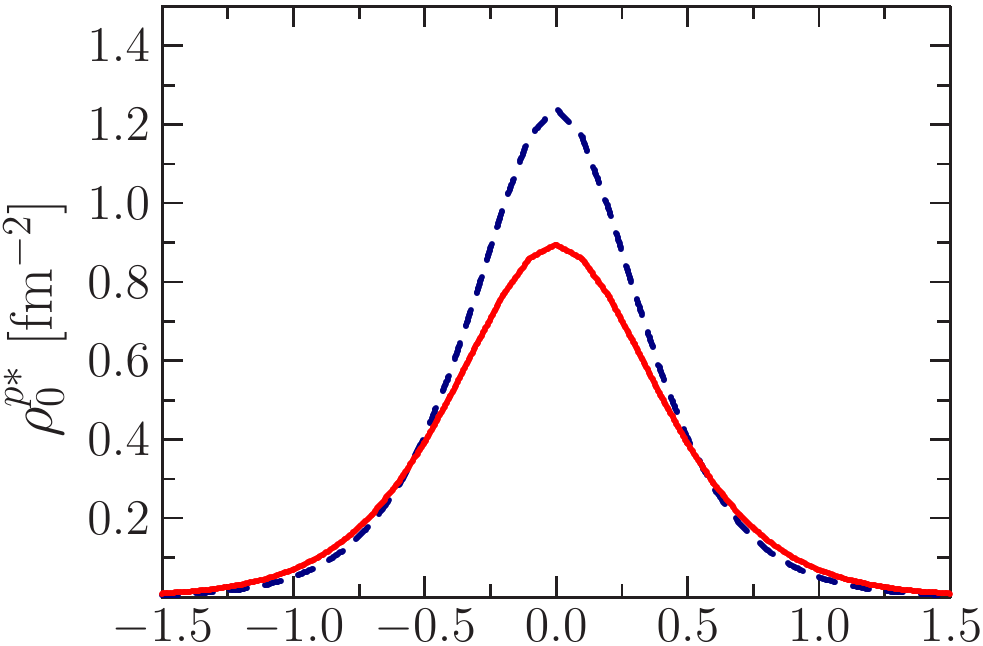}}
\hskip 0.6cm\resizebox{0.9\columnwidth}{!}{\includegraphics{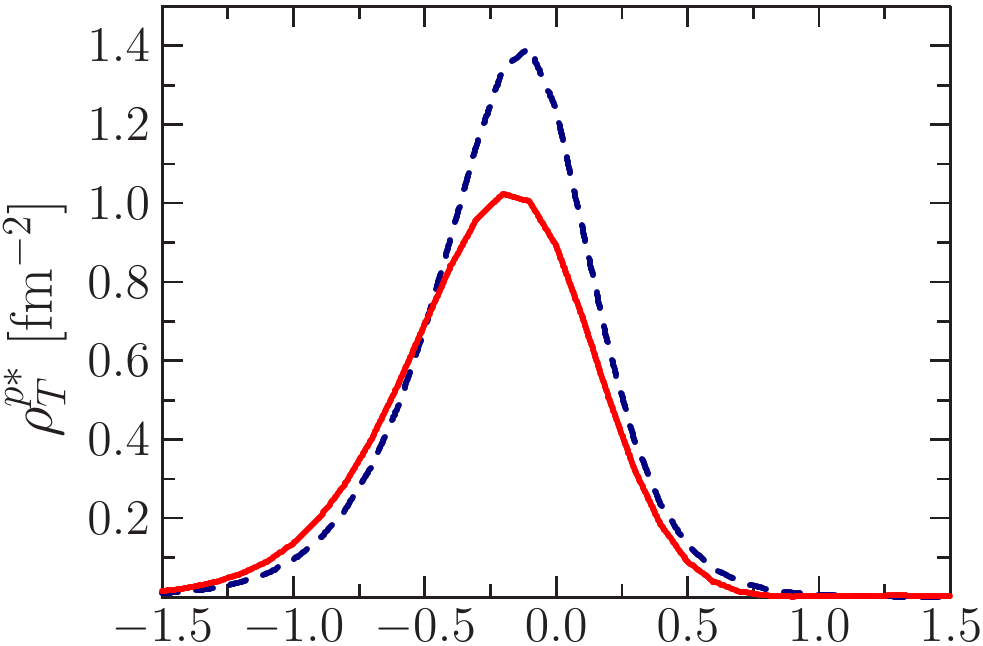}}}
\vskip 0.5cm
\centerline{\resizebox{0.9\columnwidth}{!}{\includegraphics{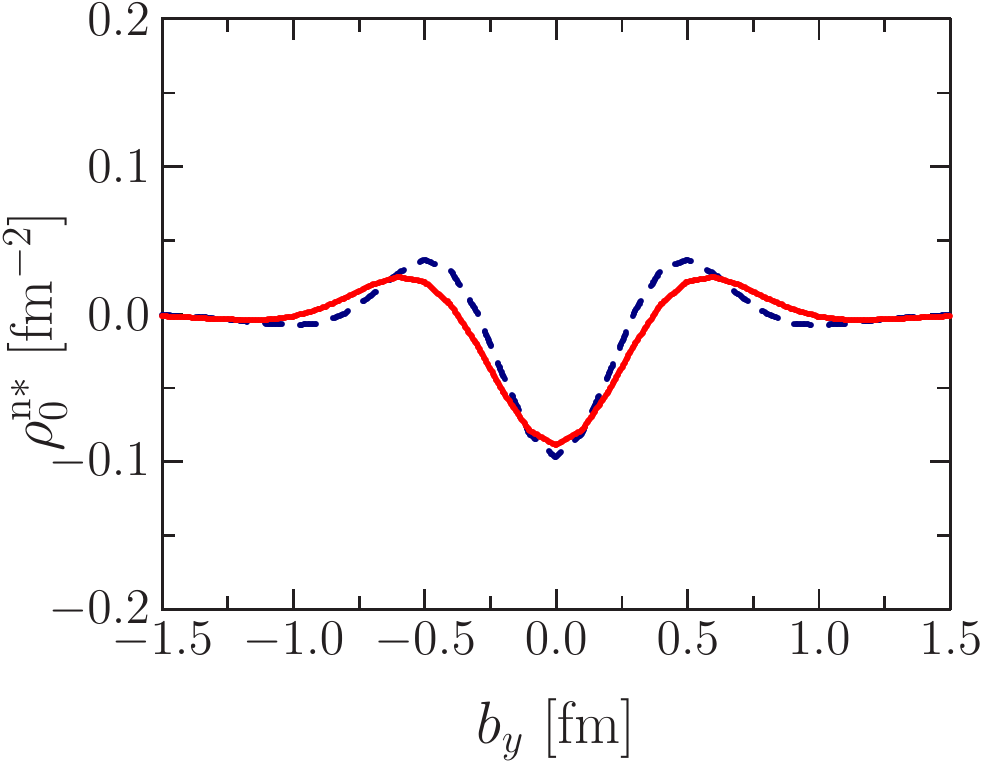}}
\hskip 0.6cm\resizebox{0.9\columnwidth}{!}{\includegraphics{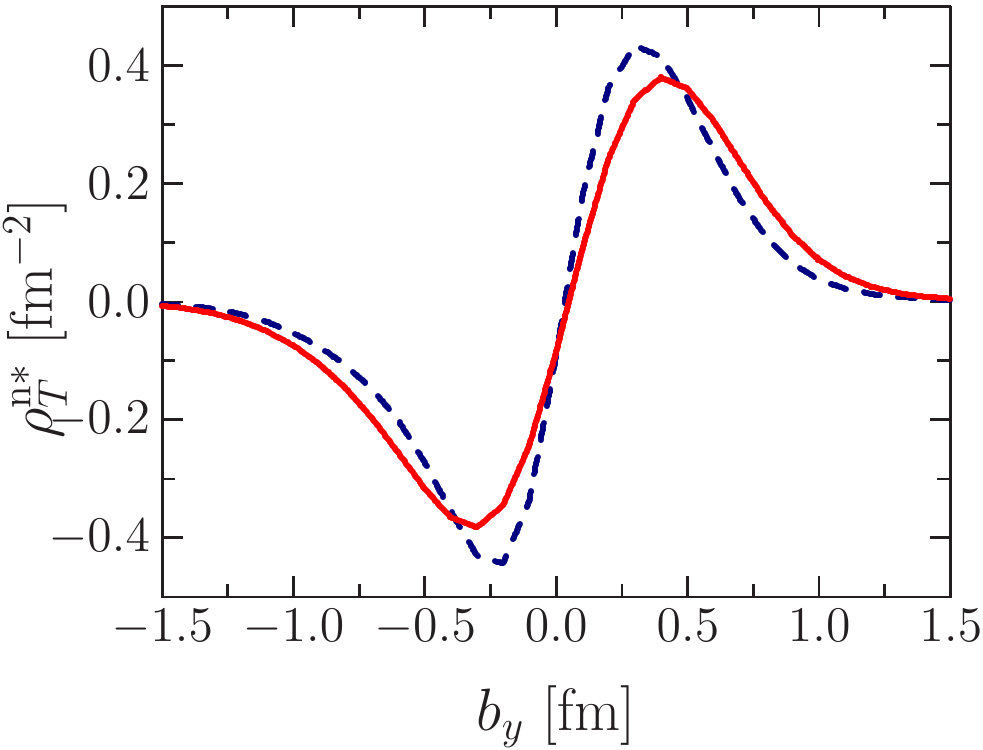}}}
\caption{Transverse charge densities inside unpolarized and
polarized protons with $b_x$ fixed to be zero are plotted in the
upper-left and the upper-right panels, respectively. In the lower
panels, those inside neutrons are presented with the same notation as
the proton case. The dashed curves draw the results in free space and
the solid ones represent those in nuclear matter
at normal nuclear matter density $\rho_0=0.5m_\pi^3$.}
\label{fig:npfixedx}
%            \hfill
\end{figure*}

When the nucleon is unpolarized, the quark transverse charge density
can be expressed as
\begin{equation}
\rho_0^*(b)\;=\; \int_0^\infty\frac{QdQ}{2\pi}\,J_0(bQ)
\frac{G_E^*(Q^2) + \tau G_M^* (Q^2)}{1+\tau},
\label{eq:trans_unpol}
\end{equation}
where $\tau=Q^2/4m_N^2$ and $J_0$ is a Bessel function of order
zero.
Figure~\ref{fig:npfixedx} depicts the transverse charge densities
inside an unpolarized proton in the upper panel and an unpolarized
neutron in the lower panel. In the left and right panels, we show
those in free space and in nuclear matter, respectively. The results
indicate that the nucleons swell up indeed in nuclear medium. In the
center of the proton, the positive charge distribution tends to
lessens but it extends in outer directions in nuclear matter. In the
case of the neutron, the medium effects are more noticeably observed
as shown in the lower-right panel of Fig.~\ref{fig:nptcm}.
\begin{figure*}[hbt]
%            \parbox{\halftext}{%   %\def\halftext{.471\textwidth}
\centerline{\resizebox{0.9\columnwidth}{!}{\includegraphics{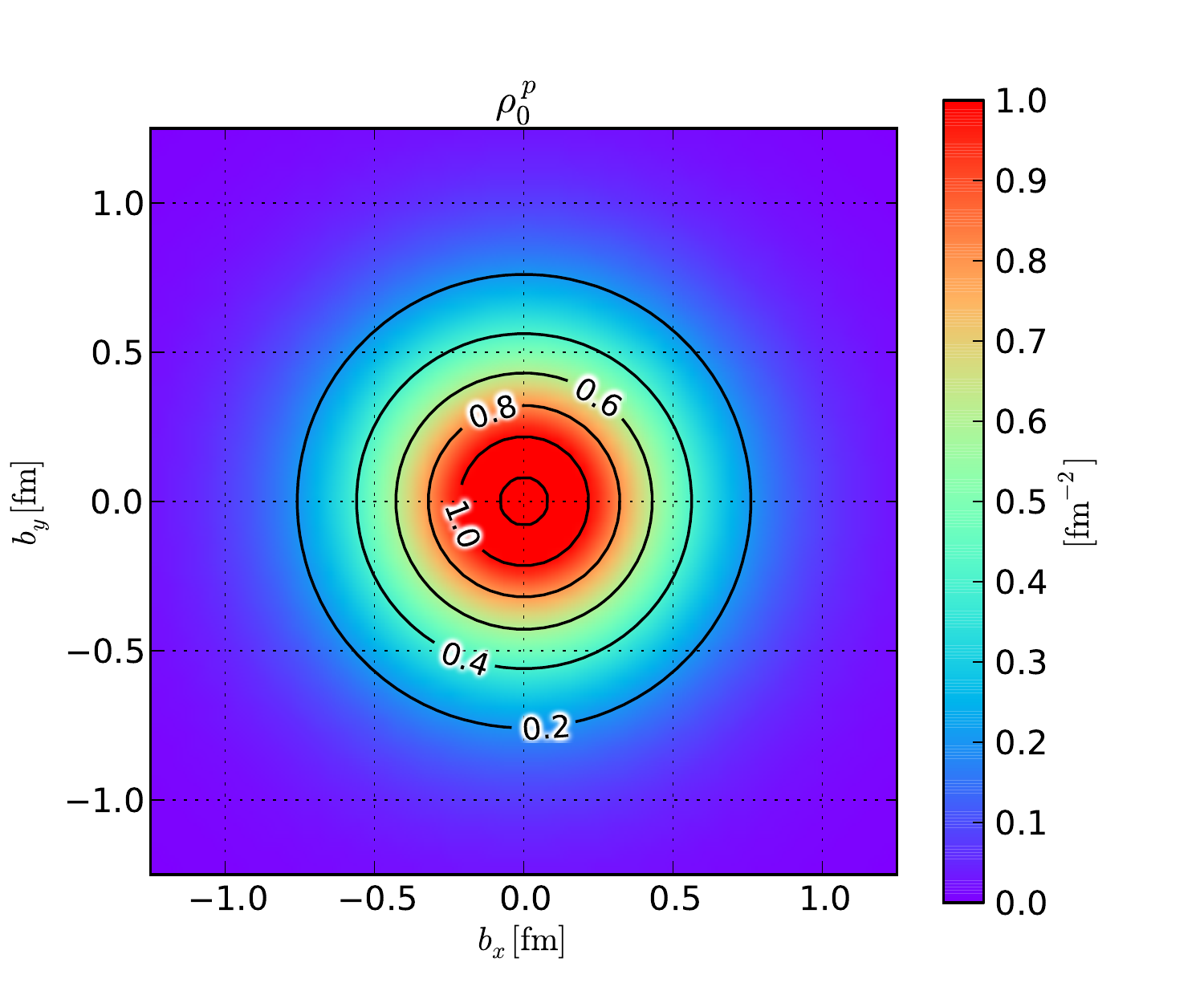}}
\hskip 0.6cm\resizebox{0.9\columnwidth}{!}{\includegraphics{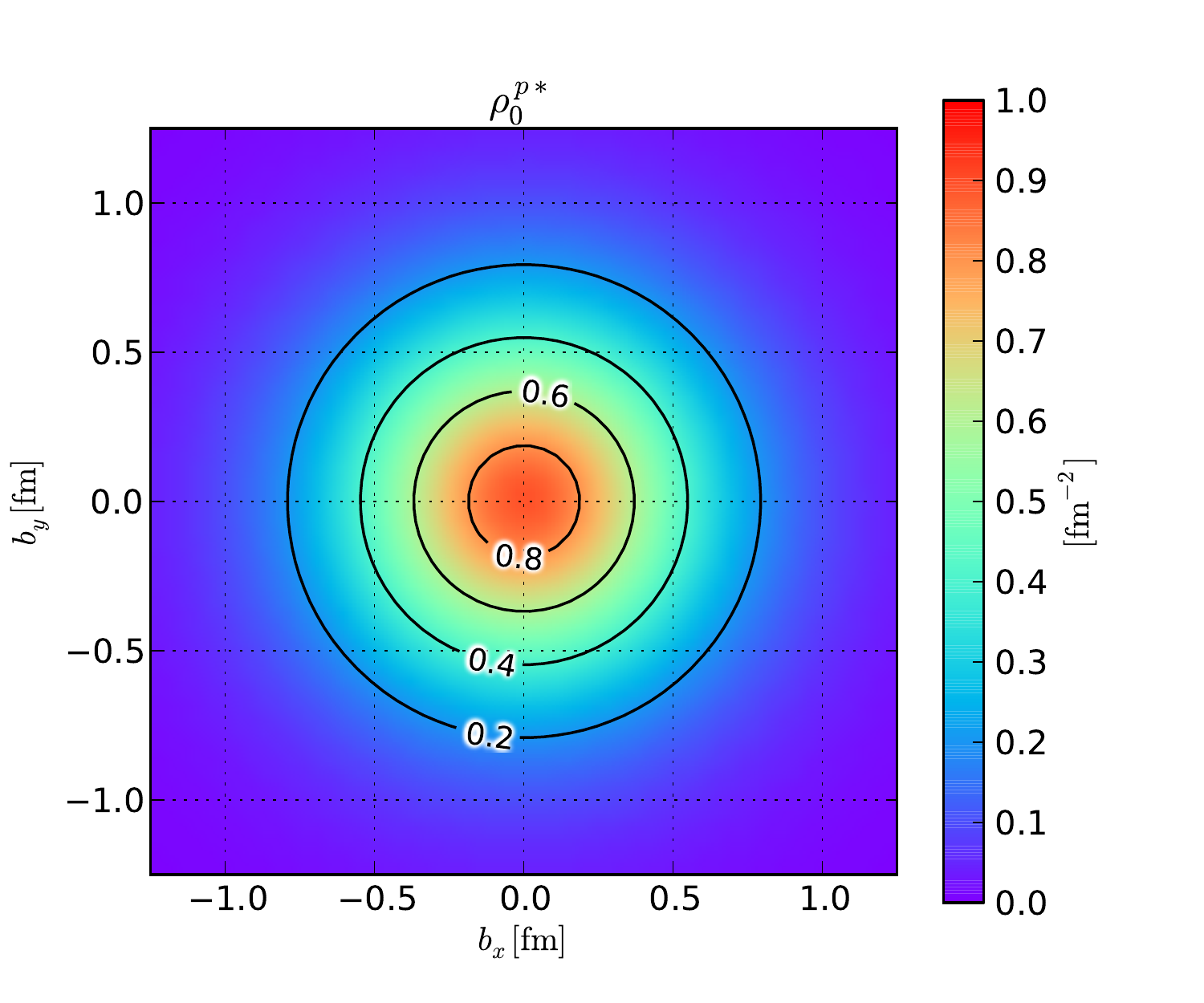}}}
\centerline{\resizebox{0.9\columnwidth}{!}{\includegraphics{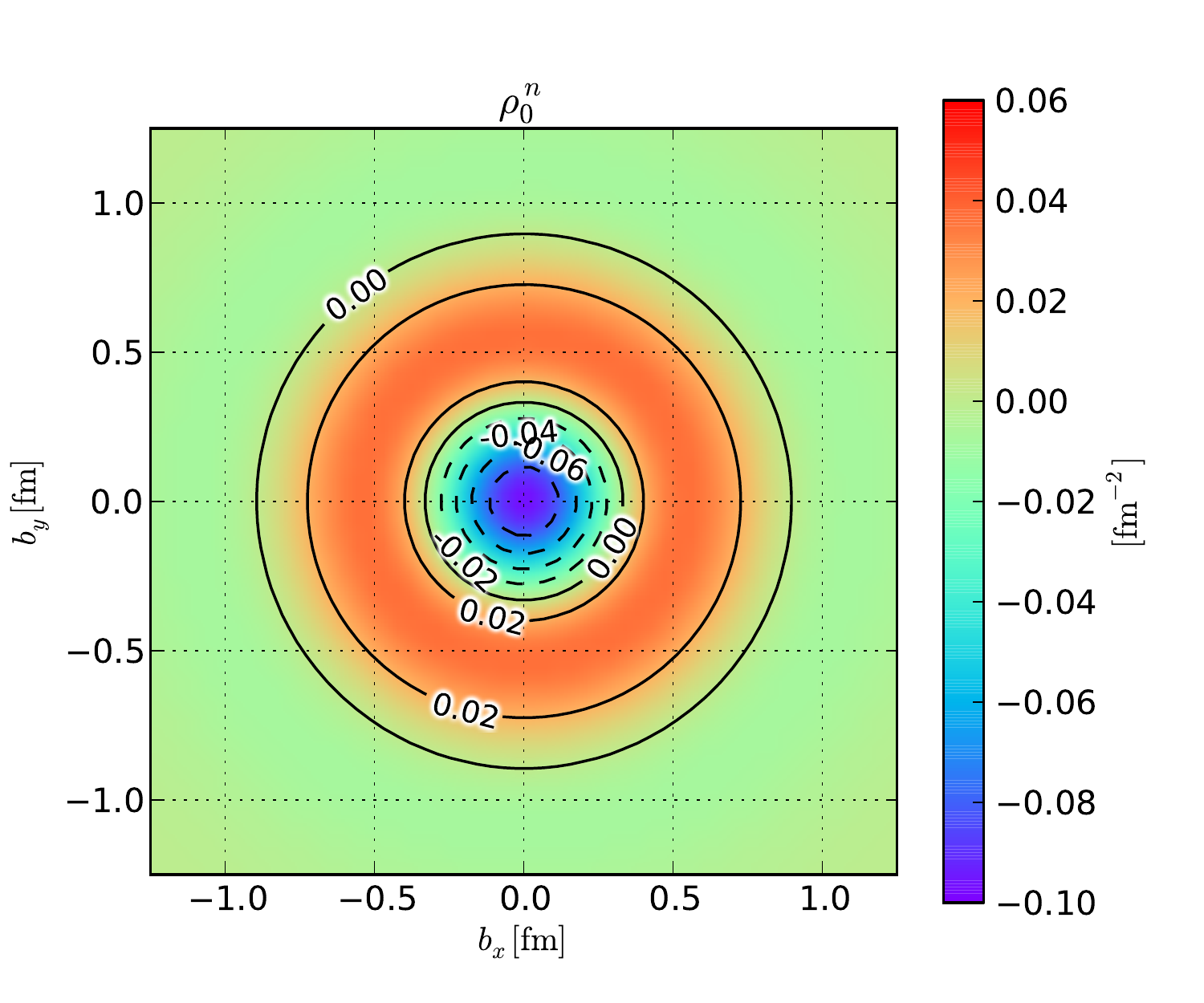}}
\hskip 0.6cm\resizebox{0.9\columnwidth}{!}{\includegraphics{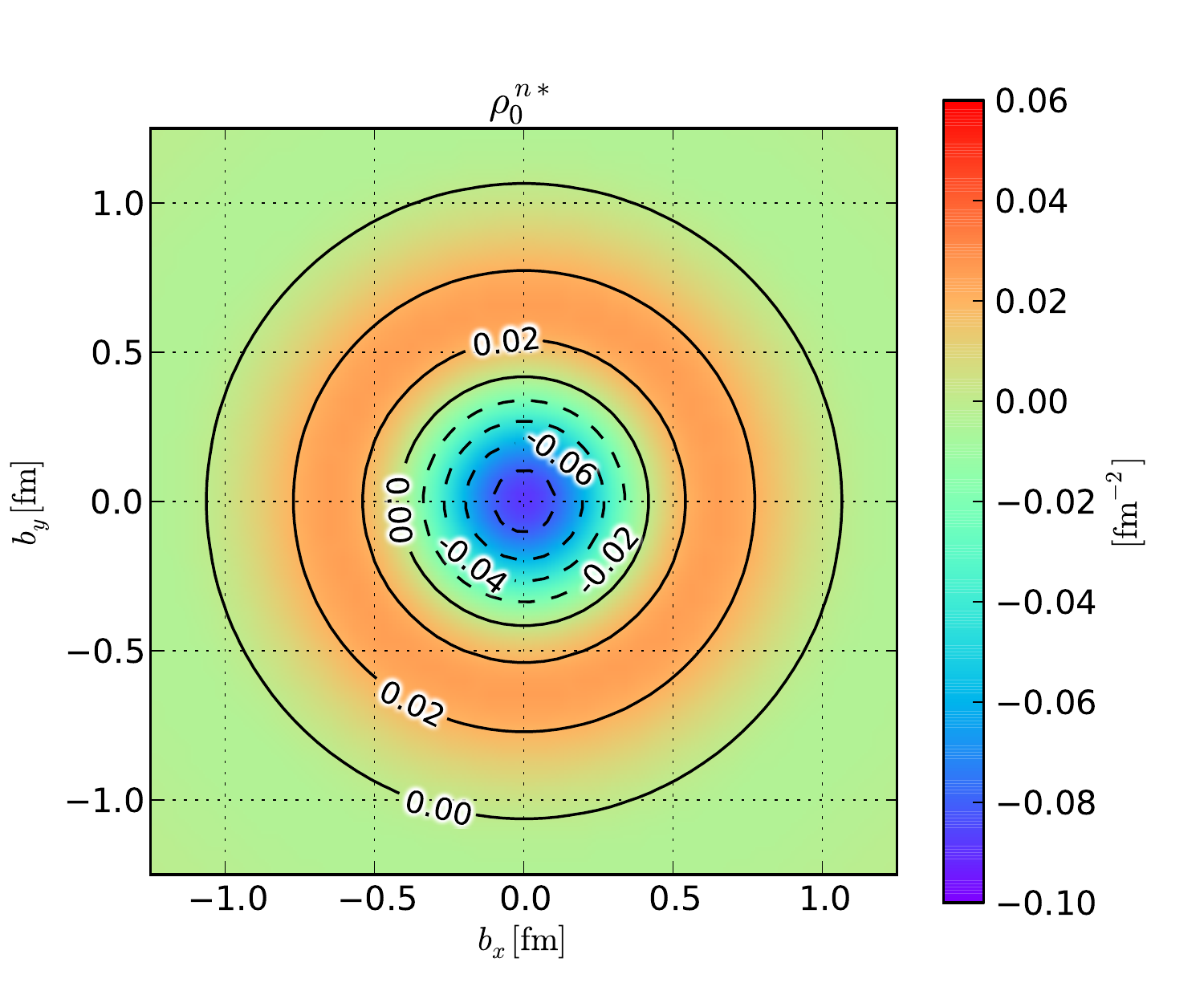}}}
\caption{Quark transverse charge densities inside an unpolarized
  proton (upper panels) and a neutron (lower panels) in free space (left
  panels) and at nuclear matter density $\rho_0=0.5m_\pi^3$ (right
  panels).}
\label{fig:nptcm}
\end{figure*}
Note, however, that the negative charges are centered as in
Ref.~\cite{Miller:2007uy} but are not as deep as that of
Ref.~\cite{Miller:2007uy}. The reason lies in the fact that the
neutron electric form factor turns out to be overestimated, as
mentioned already.

\begin{figure*}[hbt]
%            \parbox{\halftext}{%   %\def\halftext{.471\textwidth}
\centerline{\resizebox{0.9\columnwidth}{!}{\includegraphics{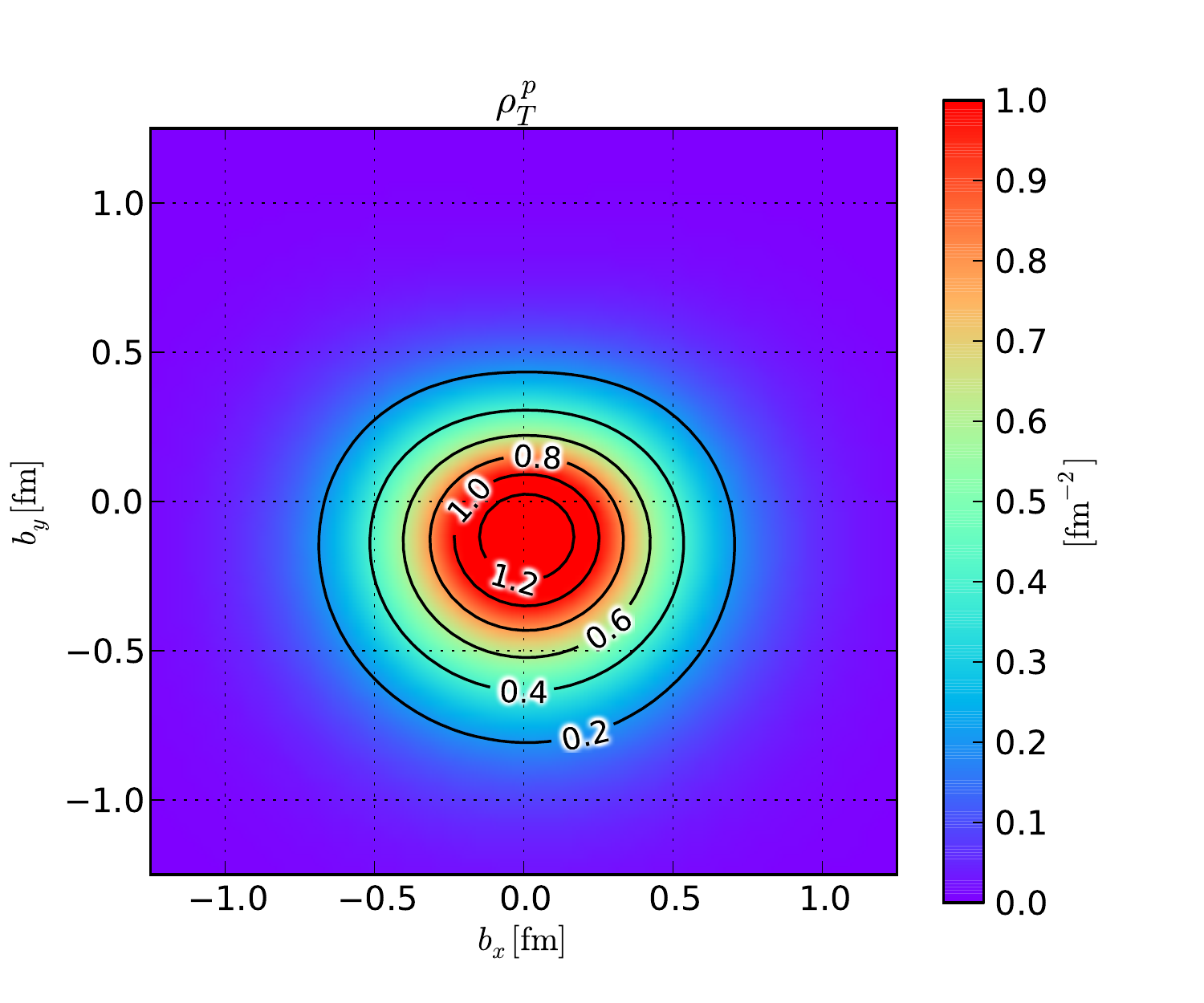}}
\hskip 0.6cm\resizebox{0.9\columnwidth}{!}{\includegraphics{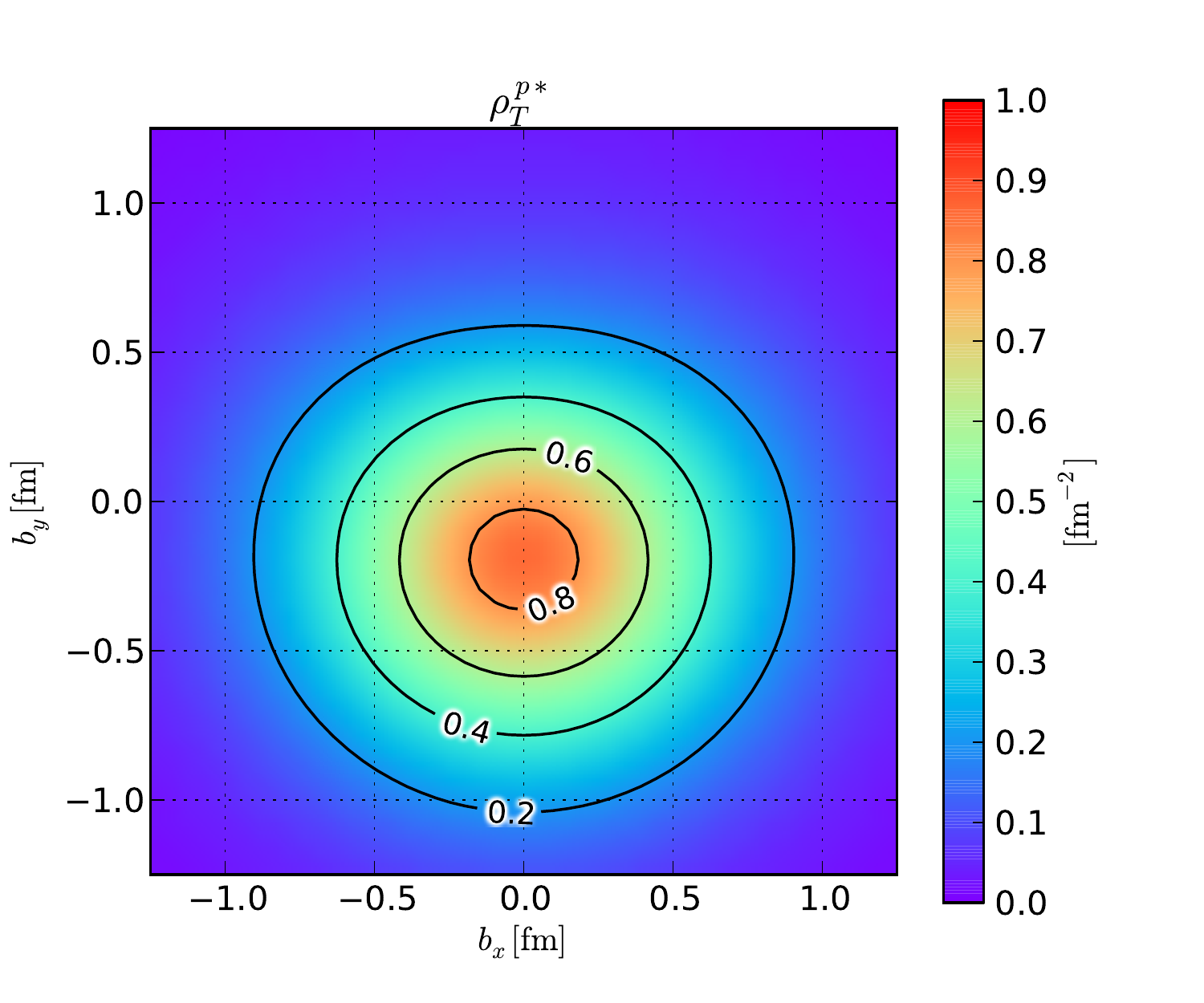}}}
\centerline{\resizebox{0.9\columnwidth}{!}{\includegraphics{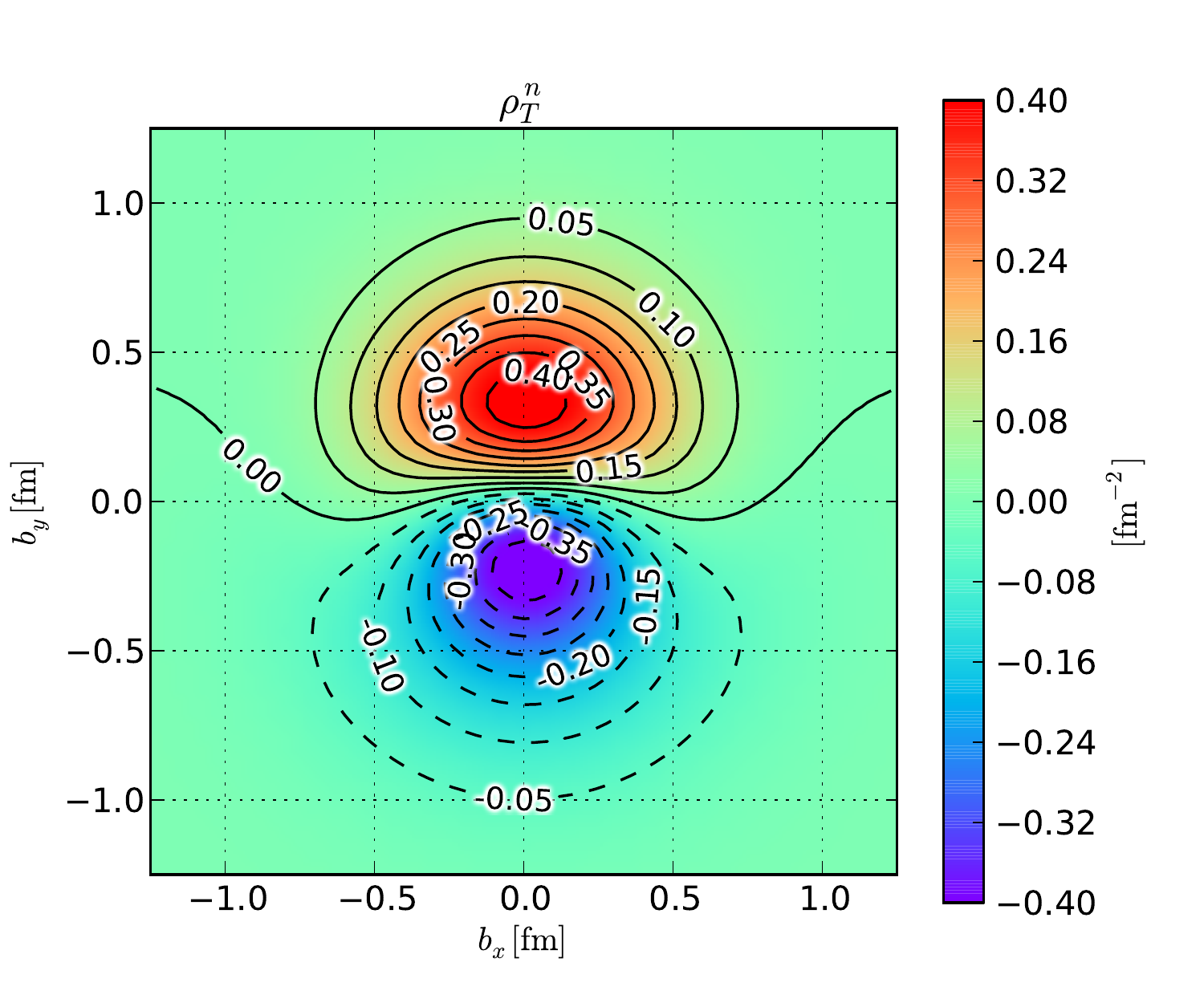}}
\hskip 0.6cm\resizebox{0.9\columnwidth}{!}{\includegraphics{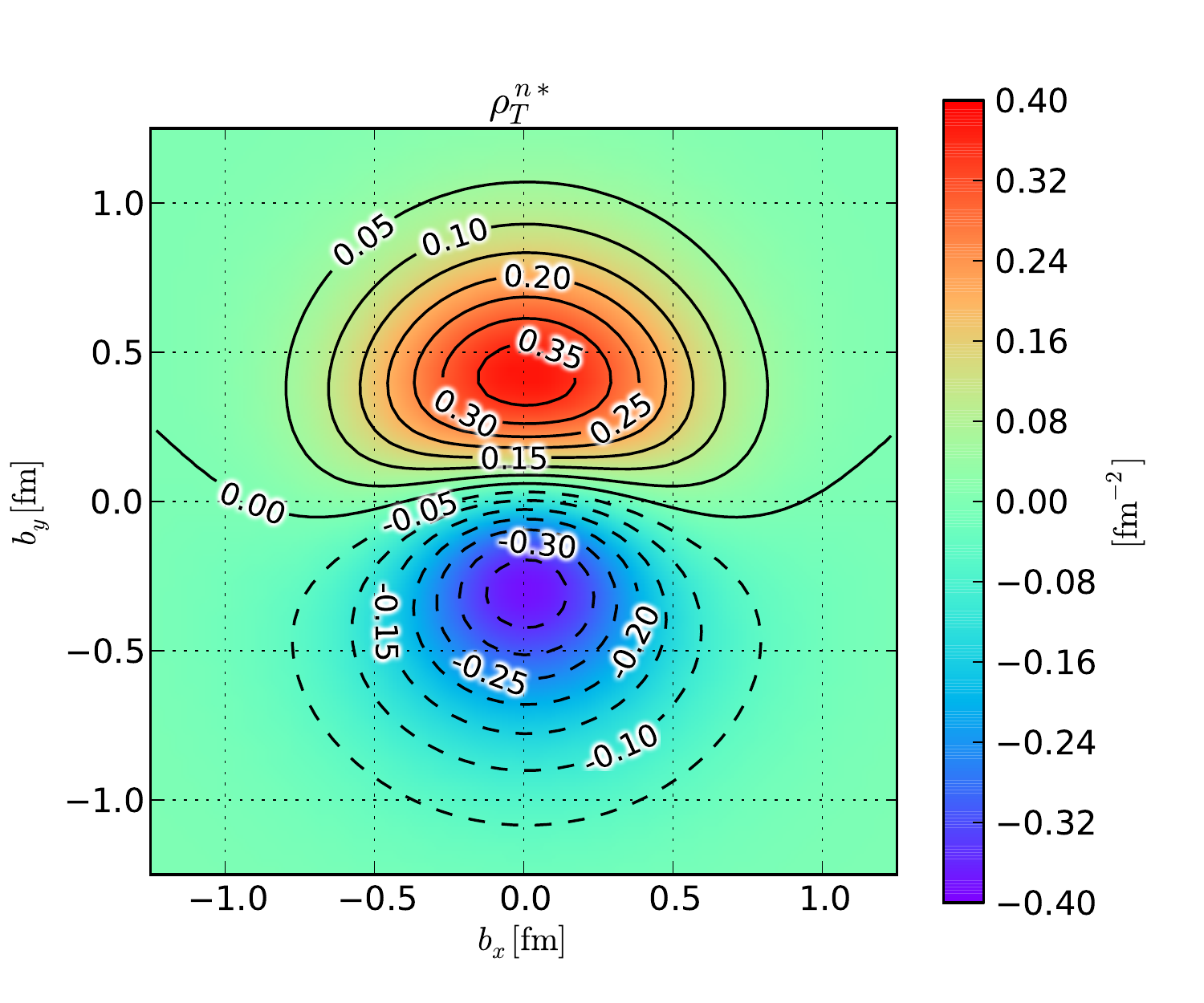}}}
\caption{Transverse charge densities of the proton (upper panels) and
neutron (lower panels) in free space (left panels) and in nuclear matter
with the density $\rho_0=0.5m_\pi^3$ (right panels).}
\label{fig:nptcm}
%            \hfill
\end{figure*}
Assuming that the nucleon is polarized in the $xy$ direction which
can be expressed by the transverse spin operator of the nucleon $\bm
S_\perp =\cos\phi_S\hat{\bm e}_x +\sin\phi_S \hat{\bm e}_y$, we find
that the Pauli form factor contributes to the transverse quark charge
densities inside a transversely polarized nucleon as follows:
\begin{eqnarray}
\rho_T^{*}(\bm b)&=&\rho_0^{*}(b)-
\sin(\phi_b-\phi_S)\cr
&&\hspace{-1cm}\times \int_0^\infty\frac{Q^2dQ}{4\pi m_N}
J_1(bQ)\frac{-G_E^*(Q^2) + G_M^*(Q^2)}{1+\tau}\,,
\label{trchden}
\end{eqnarray}
where $\bm b=b(\cos\phi_b\hat{\bm   e}_x+\sin\phi_b\hat{\bm e}_y)$
denotes the position vector from the center of the nucleon in the
transverse plane. The $J_{0,1}$ are the Bessel functions of order 0
and 1. The second term in Eq.~(\ref{trchden}) makes the transverse
charge densities distorted from the unpolarized one. We choose the
polarization direction of the nucleon along the $x$ axis,
i.e. $\phi_S=0$.

In Fig.~\ref{fig:npfixedx}, we draw the transverse charge densities
inside polarized nucleons. The density inside the polarized nucleon is
shown to be distorted in the negative $b_y$ direction.
As explained already in Refs.~\cite{Burkardt:2002hr,Carlson:2007xd},
the polarization of the nucleon in the $x$ direction induces the
electric dipole moment in the negative $y$ direction, which is just
the relativistice effect. As a result, the strong distortion arises as
shown in Fig.~~\ref{fig:npfixedx}.
In nuclear matter, the transverse charge density shows the same
tendency but seems less obvious. This might be due to the fact that
the quark charges decrease in the center of the proton but extend to
the outer region. Hence, the influence of the proton polarization on
the transverse charge densities is somewhat diminished.

When it comes to the neutron case, the polarization effects are
profound: the negative charges are forced to move to the negative
$y$, while the positive charges are found in the positive $y$. As
a result, the transverse charge densities inside the polarized neutron
turn out be very asymmetric. The same is true in nuclear matter.
In Fig.~\ref{fig:npfixedx}, the profiles of the transverse charge
densities are plotted with $b_x$ is fixed to be zero. The effects of
nuclear medium are clearly observed. In general, the transverse charge
densities in nuclear matter are likely to decrease in the center but
extends to the outer directions.

\vspace{0.5cm}
%\section{Summary}
\textbf{5.} The present work aimed at investigating how the
electromagnetic properties of the nucleons were modified in nuclear
matter within the framework of the in-medium modified Skyrme
model. We found that the electromagnetic form factors of the nucleon
fell off faster in nuclear matter. The transverse charge
densities inside a proton and a neutron indicate that the nucleon
in nuclear medium tends to swell up in general. The effects of the
proton polarization are somewhat reduced in nuclear matter. Since the
positive charges decrease in the center of the proton but extends to
outer directions, the distortion of the transverse charge distribution
is less noticeable. The same features were found in the neutron case.

The present results are consistent with those of other models: for
example, a soliton model with vector mesons~\cite{Meissner:1988wj} and
the QMC model calculations~\cite{Lu:1997mu}. The present investigation
may shed light on understanding the medium modification of the
nucleon, which will be measured experimentally in near future.
\vspace{0.5cm}

%---------------------------------
\section*{Acknowledgments}
%---------------------------------
H.Ch.K is grateful to P. Navratil and R. Woloshyn at TRIUMF for their
hospitality during his visit, where the present work was initiated.
The authors are thankful to J.-H. Jung for preparing the figures for
the transverse charge densities. The work is supported by Basic
Science Research Program through the
National Research Foundation of Korea (NRF) funded by
the Ministry of Education, Science and Technology Grant Number:
2011-0023478 (U.Y.) and Grant Number: 2012004024 (H.Ch.K.).


\begin{thebibliography}{99}

%\rcite{Jones:1999rz}
\bibitem{Jones:1999rz}
  M.~K.~Jones {\it et al.}  [Jefferson Lab Hall A Collaboration],
  %``G(E(p)) / G(M(p)) ratio by polarization transfer in polarized e p
  %---> e polarized p,''
  Phys.\ Rev.\ Lett.\  {\bf 84} (2000) 1398
  [nucl-ex/9910005].
  %%CITATION = NUCL-EX/9910005;%%

%\rcite{Gayou:2001qt}
\bibitem{Gayou:2001qt}
  O.~Gayou {\it et al.}  [Jefferson Lab Hall A Collaboration],
  %``Measurements of the elastic electromagnetic form-factor ratio
  %mu(p) G(Ep) / G(Mp) via polarization transfer,''
  Phys.\ Rev.\ C {\bf 64} (2001) 038202.
  %%CITATION = PHRVA,C64,038202;%%

%\rcite{Gayou:2001qd}
\bibitem{Gayou:2001qd}
  O.~Gayou {\it et al.}  [Jefferson Lab Hall A Collaboration],
  %``Measurement of G(Ep) / G(Mp) in polarized-e p ---> e polarized-p
  %to Q**2 = 5.6-GeV**2,''
  Phys.\ Rev.\ Lett.\  {\bf 88} (2002) 092301
  [nucl-ex/0111010].
  %%CITATION = NUCL-EX/0111010;%%

%\rcite{Punjabi:2005wq}
\bibitem{Punjabi:2005wq}
  V.~Punjabi {\it et al.}  [Jefferson Lab Hall A Collaboration],
  %``Proton elastic form-factor ratios to Q**2 = 3.5-GeV**2 by
  %polarization transfer,''
  Phys.\ Rev.\ C {\bf 71} (2005) 055202
  [Erratum-ibid.\ C {\bf 71} (2005) 069902]
  [nucl-ex/0501018].
  %%CITATION = NUCL-EX/0501018;%%

%\rcite{Puckett:2010ac}
\bibitem{Puckett:2010ac}
  A.~J.~R.~Puckett, E.~J.~Brash, M.~K.~Jones, W.~Luo, M.~Meziane,
  L.~Pentchev, C.~F.~Perdrisat and V.~Punjabi {\it et al.},
  %``Recoil Polarization Measurements of the Proton Electromagnetic
  %Form Factor Ratio to Q^2 = 8.5 GeV^2,''
  Phys.\ Rev.\ Lett.\  {\bf 104} (2010) 242301
  [arXiv:1005.3419 [nucl-ex]].
  %%CITATION = ARXIV:1005.3419;%%

%\rcite{Bernauer:2010wm}
\bibitem{Bernauer:2010wm}
  J.~C.~Bernauer {\it et al.}  [A1 Collaboration],
  %``High-precision determination of the electric and magnetic form
  %factors of the proton,''
  Phys.\ Rev.\ Lett.\  {\bf 105} (2010) 242001
  [arXiv:1007.5076 [nucl-ex]].
  %%CITATION = ARXIV:1007.5076;%%

%\rcite{Ron:2011rd}
\bibitem{Ron:2011rd}
  G.~Ron {\it et al.}  [Jefferson Lab Hall A Collaboration],
  %``Low $Q^2$ measurements of the proton form factor ratio $mu_p G_E
  %/ G_M$,''
  Phys.\ Rev.\ C {\bf 84} (2011) 055204
  [arXiv:1103.5784 [nucl-ex]].
  %%CITATION = ARXIV:1103.5784;%%

%\rcite{Zhan:2011ji}
\bibitem{Zhan:2011ji}
  X.~Zhan, K.~Allada, D.~S.~Armstrong, J.~Arrington, W.~Bertozzi,
  W.~Boeglin, J.~-P.~Chen and K.~Chirapatpimol {\it et al.},
  %``High Precision Measurement of the Proton Elastic Form Factor
  %Ratio $\mu_pG_E/G_M$ at low $Q^2$,''
  Phys.\ Lett.\ B {\bf 705} (2011) 59
  [arXiv:1102.0318 [nucl-ex]].
  %%CITATION = ARXIV:1102.0318;%%

%\rcite{HydeWright:2004gh}
\bibitem{HydeWright:2004gh}
  C.~E.~Hyde-Wright and K.~de Jager,
  %``Electromagnetic form factors of the nucleon and Compton scattering,''
  Ann.\ Rev.\ Nucl.\ Part.\ Sci.\  {\bf 54} (2004) 217
  [nucl-ex/0507001].
  %%CITATION = NUCL-EX/0507001;%%

%\rcite{Arrington:2006zm}
\bibitem{Arrington:2006zm}
  J.~Arrington, C.~D.~Roberts and J.~M.~Zanotti,
  %``Nucleon electromagnetic form-factors,''
  J.\ Phys.\ G {\bf 34} (2007) S23
  [nucl-th/0611050].
  %%CITATION = NUCL-TH/0611050;%%

%\rcite{Perdrisat:2006hj}
\bibitem{Perdrisat:2006hj}
  C.~F.~Perdrisat, V.~Punjabi and M.~Vanderhaeghen,
  %``Nucleon Electromagnetic Form Factors,''
  Prog.\ Part.\ Nucl.\ Phys.\  {\bf 59} (2007) 694
  [hep-ph/0612014].
  %%CITATION = HEP-PH/0612014;%%

%\rcite{Cates:2011pz}
\bibitem{Cates:2011pz}
  G.~D.~Cates, C.~W.~de Jager, S.~Riordan and B.~Wojtsekhowski,
  %``Flavor decomposition of the elastic nucleon electromagnetic form
  %factors,''
  Phys.\ Rev.\ Lett.\  {\bf 106}, 252003 (2011)
  [arXiv:1103.1808 [nucl-ex]].
  %%CITATION = ARXIV:1103.1808;%%

%\rcite{Qattan:2012zf}
\bibitem{Qattan:2012zf}
  I.~A.~Qattan and J.~Arrington,
  %``Flavor decomposition of the nucleon electromagnetic form
  %factors,''
  Phys.\ Rev.\ C {\bf 86}, 065210 (2012)
  [arXiv:1209.0683 [nucl-ex]].
  %%CITATION = ARXIV:1209.0683;%%

%\cite{Burkardt:2002hr}
\bibitem{Burkardt:2002hr}
  M.~Burkardt,
  %``Impact parameter space interpretation for generalized parton
  %distributions,''
Int.\ J.\ Mod.\ Phys.\ A \textbf{18} (2003) 173 [hep-ph/0207047].
%%CITATION = HEP-PH/0207047;%%  %340 citations
  %counted in INSPIRE as of 30 Mar 2013

%\cite{Burkardt:2000za}
\bibitem{Burkardt:2000za}
  M.~Burkardt,
  %``Impact parameter dependent parton distributions and off forward
  %parton distributions for zeta ---> 0,''
Phys.\ Rev.\ D {\bf 62} (2000) 071503   [Erratum-ibid.\ D {\bf 66}
(2002) 119903]  [hep-ph/0005108].
%%CITATION = HEP-PH/0005108;%%  %371 citations counted in INSPIRE as
%%of 30 Mar 2013

%\rcite{Miller:2007uy}
\bibitem{Miller:2007uy}
  G.~A.~Miller,
  %``Charge Density of the Neutron,''
  Phys.\ Rev.\ Lett.\  {\bf 99} (2007) 112001
  [arXiv:0705.2409 [nucl-th]].
  %%CITATION = ARXIV:0705.2409;%%

%\rcite{Carlson:2007xd}
\bibitem{Carlson:2007xd}
  C.~E.~Carlson and M.~Vanderhaeghen,
  %``Empirical transverse charge densities in the nucleon and the
  %nucleon-to-Delta transition,''
  Phys.\ Rev.\ Lett.\  {\bf 100} (2008) 032004
  [arXiv:0710.0835 [hep-ph]].
  %%CITATION = ARXIV:0710.0835;%%
%\cite{Airapetian:2009bi}
\bibitem{Airapetian:2009bi}
  A.~Airapetian {\it et al.}  [HERMES Collaboration],
  %``Nuclear-mass dependence of azimuthal beam-helicity and
  %beam-charge asymmetries in deeply virtual Compton scattering,''
Phys.\ Rev.\ C {\bf 81} (2010) 035202  [arXiv:0911.0091 [hep-ex]].
  %%%CITATION = ARXIV:0911.0091;%%  %21 citations counted in INSPIRE
  %%%as of 30 Mar 2013

\bibitem{Rakhimov:1996vq}
  A.~Rakhimov, M.~M.~Musakhanov, F.~C.~Khanna and U.~T.~Yakhshiev,
  %``Medium modification of nucleon properties in Skyrme model,''
  Phys.\ Rev.\ C {\bf 58} (1998) 1738 [nucl-th/9609049].
  %%CITATION = NUCL-TH/9609049;%%

\bibitem{Yakhshiev:2010kf}
  U.~Yakhshiev and H.-Ch.~Kim,
  %``Binding energy per nucleon and hadron properties in nuclear matter,''
  Phys. Rev.  C {\bf 83} (2011) 038203 [arXiv:1009.2909 [hep-ph]].
  %%CITATION = ARXIV:1009.2909;%%

%\cite{Kim:2012ts}
\bibitem{Kim:2012ts}
  H.~-Ch.~Kim, P.~Schweitzer, and U.~Yakhshiev,
  %``Energy-momentum tensor form factors of the nucleon in nuclear
  %matter,''
Phys.\ Lett.\ B {\bf 718} (2012) 625  [arXiv:1205.5228 [hep-ph]].
%% CITATION = ARXIV:1205.5228;%%  %3 citations counted in INSPIRE as
%% of 30 Mar 2013

\bibitem{Ericsonbook}
  T.~Ericson and W.~Weise, {\em Pions and Nuclei} (Clarendon, Oxford,
  1988).
\bibitem{Lu:1997mu}
  D.~H.~Lu, A.~W.~Thomas, K.~Tsushima, A.~G.~Williams and K.~Saito,
  %``In-medium electron - nucleon scattering,''
  Phys.\ Lett.\  B {\bf 417} (1998) 217.
%  [arXiv:nucl-th/9706043].
  %%CITATION = PHLTA,B417,217;%%
\bibitem{Meissner:1988wj}
  U.~G.~Meissner,
  %``MEDIUM MODIFICATIONS OF THE NEUTRON CHARGE FORM-FACTOR,''
  Phys.\ Rev.\ Lett.\  {\bf 62} (1989) 1013.
  %%CITATION = PRLTA,62,1013;%%

\bibitem{Miller:1990iz}
  G.~A.~Miller, B.~M.~K.~Nefkens, I.~Slaus,
  %``Charge symmetry, quarks and mesons,''
Phys.\ Rept.\  {\bf 194} (1990) 1.  %%CITATION = PRPLC,194,1;%%
\bibitem{Beck:2001yx}
  D.~H.~Beck, R.~D.~McKeown,
  %``Parity violating electron scattering and nucleon structure,''
  Ann.\ Rev.\ Nucl.\ Part.\ Sci.\  {\bf 51} (2001) 189
  [hep-ph/0102334].  %%CITATION = HEP-PH/0102334;%%

%\cite{Kelly:2002if}
\bibitem{Kelly:2002if}
  J.~J.~Kelly,
  %``Nucleon charge and magnetization densities from Sachs
  %form-factors,''
Phys.\ Rev.\ C {\bf 66} (2002) 065203  [hep-ph/0204239].
 %%CITATION =HEP-PH/0204239;%%
%73 citations counted in INSPIRE as of 12 Apr 2013
\end{thebibliography}
\end{document}